\documentclass[traditabstract]{aa}
\usepackage{txfonts}

\usepackage{natbib}
\bibpunct{(}{)}{;}{a}{}{,} 

\usepackage{graphicx}
\usepackage{ amssymb }
\usepackage[hyphens]{url}

\begin{document}

\title{Absence of hot gas within the Wolf-Rayet bubble around
  WR\,16} \author{J.A.\,Toal\'{a} \and M.A.\,Guerrero}

\institute{Instituto de Astrof\'\i sica de Andaluc\'\i a,
  IAA-CSIC, Glorieta de la Astronom\'\i a s/n, 18008 Granada, 
Spain; toala@iaa.es}

\abstract{We present the analysis of \emph{XMM-Newton} archival
  observations towards the Wolf-Rayet (WR) bubble around WR\,16.
  Despite the closed bubble morphology of this WR nebula, the
  \emph{XMM-Newton} observations show no evidence of diffuse emission
  in its interior as in the similar WR bubbles NGC\,6888 and S\,308.
  We use the present observations to estimate a 3-$\sigma$ upper limit
  to the X-ray luminosity in the 0.3--1.5~keV energy band equal to
  7.4$\times$10$^{32}$~erg~s$^{-1}$ for the diffuse emission from the
  WR nebula, assuming a distance of 2.37~kpc. The WR nebula around
  WR\,16 is the fourth observed by the current generation of X-ray
  satellites and the second not detected.  We also examine \emph{FUSE}
  spectra to search for nebular O~{\sc vi} absorption lines in the
  stellar continuum of WR\,16. The present far-UV data and the lack of
  measurements of the dynamics of the optical WR bubble do not allow
  us to confirm the existence of a conductive layer of gas at
  T$\sim$3$\times$10$^5$~K between the cold nebular gas and the hot
  gas in its interior. The present observations result in an upper
  limit of $n_{\mathrm{e}}<$0.6~cm$^{-3}$ on the electron density of
  the X-ray emitting material within the nebula.}


\titlerunning{Absence of hot gas within the Wolf-Rayet bubble around
  WR\,16}

\authorrunning{Toal\'a \& Guerrero}

\maketitle

\section{Introduction}

When massive stars reach the Wolf-Rayet (WR) stage in their late
evolution, powerful stellar winds carve the circumstellar medium and
produce WR nebulae. X-ray emission from these objects is expected to
arise from shock-heated plasma that is produced when the WR stellar
wind rams into material previously ejected during the red supergiant
(RSG) or luminous blue variable (LBV) phase \citep{GSML1995}.  To
date, there are only three WR nebulae reported in the literature that
had been observed with the latest generation of X-ray satellites
(\emph{Chandra}, \emph{Suzaku}, and \emph{XMM-Newton}): S\,308,
RCW\,58, and NGC\,6888 around WR\,6, WR\,40, and WR\,136, respectively
\citep{Chu2003,Gosset2005,Chu2006,Zhekov2011,Toala2012}.

The most recent observations towards S\,308 and NGC\,6888 show a
dominant X-ray-emitting plasma component with a temperature
$\gtrsim$10$^{6}$~K and abundances similar to those of the optical
nebulae \citep{Zhekov2011,Toala2012}.  The temperature and abundances
of the dominant X-ray-emitting plasma can be explained by means of
mixing between the outer cold material and the shocked fast stellar
wind.  On the other hand, RCW\,58 has not been detected by
\emph{XMM-Newton} \citep{Gosset2005}.  The cause of this non-detection
of hot gas may be twofold.  First, WR nebulae lay within the Galactic
Plane, and their soft X-ray emission may be highly affected by
absorption (RCW\,58 has a Galactic latitude $\sim$5\degr\, and the
hydrogen column towards it is 5$\times$10$^{21}$ cm$^{-2}$.).
Secondly, RCW\,58 presents a disrupted shell morphology
\citep{Gruendl2000}, which can imply that the hot gas has already
escaped outside the nebula, greatly reducing the X-ray emissivity
\citep{Toala2011}.  The limited number of X-ray observations of WR
nebulae makes it still an exploration field.  It is thus highly
important to increase the number of observed WR nebulae with good
spatial coverage, high sensitivity, and suitable energy resolution for
spectral analyses.

\begin{figure*}
\begin{center}
\includegraphics[width=0.5\linewidth]{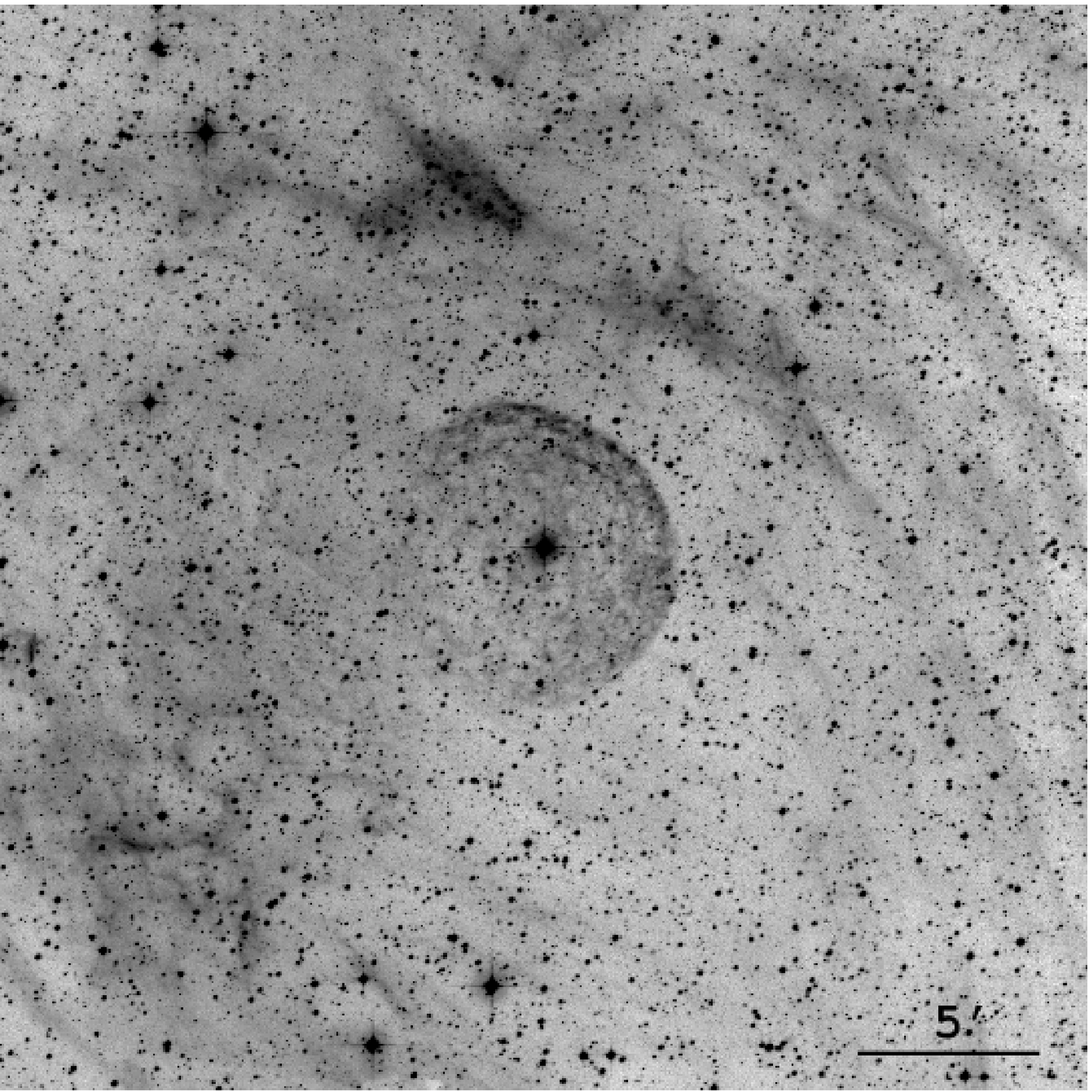}~
\includegraphics[width=0.5\linewidth]{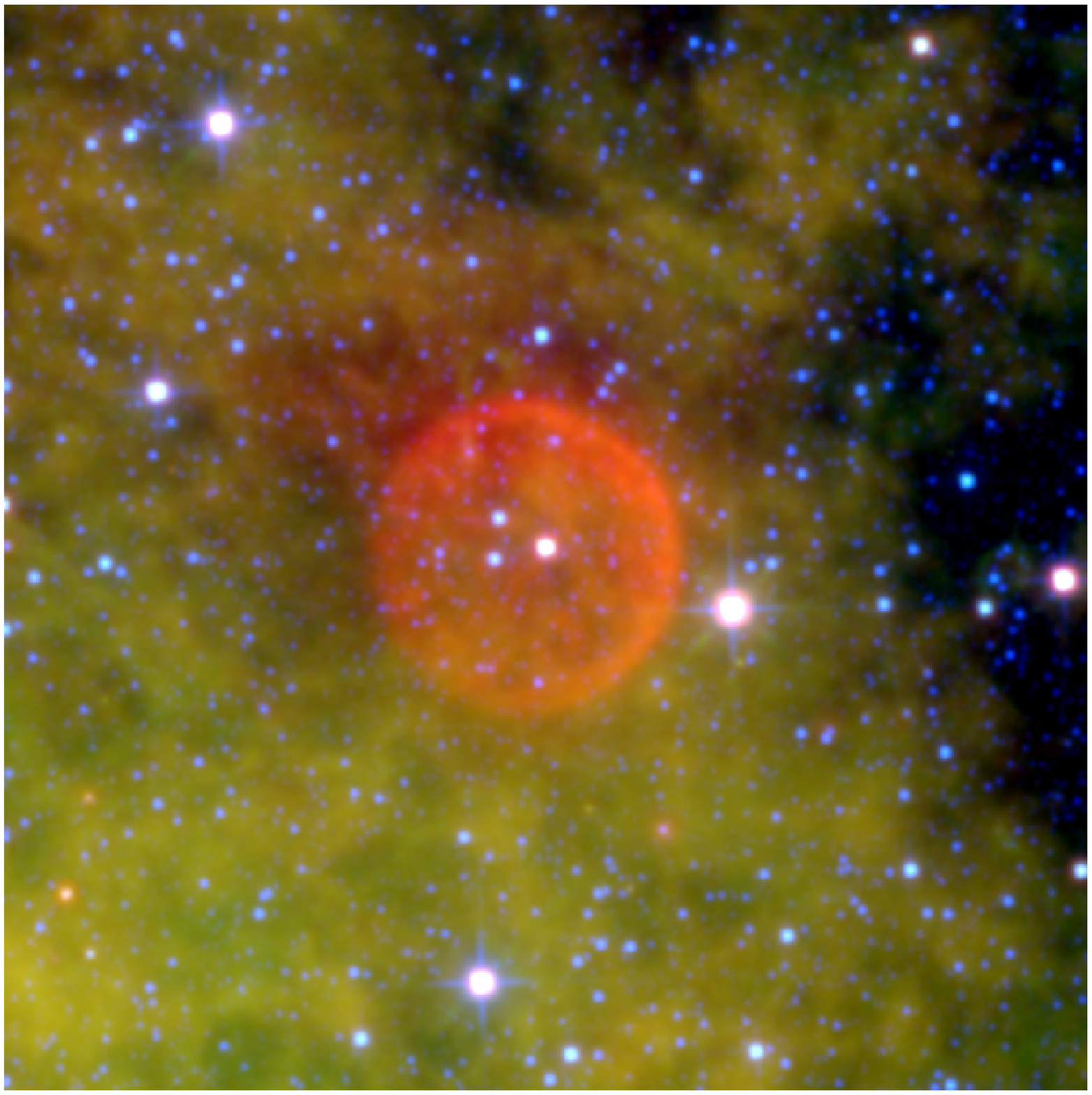}
\end{center}
\caption{
{\it (left)} 
Narrow-band H$\alpha$ image of the WR nebula around WR\,16 taken from 
the Super COSMOS Sky Survey \citep{Parker2005}. 
{\it (right)}
Color-composite mid-infrared \emph{WISE} W2 4.6~$\mu$m (blue), W3 
12~$\mu$m (green), and W4 22~$\mu$m (red) picture of the WR nebula 
around WR\,16.  
The central star, WR\,16, is located at the center of each image. 
North is up, east to the left.
}
\label{fig:imagen_wr16}
\end{figure*}

The soft nature of the X-ray emission detected in S\,308 and NGC\,6888
needs further discussion.  Theoretical models of adiabatic shocked
winds predict high plasma temperatures, $>$10$^7$~K
\citep[e.g.,][]{Dyson1997}, which are not reported by observations.
The relatively low temperature of the X-ray-emitting plasma is often
attributed to mixing in the wind-wind interaction zone, such as,
thermal conduction. The hot gas will be cooled down to temperatures of
10$^6$~K when in contact with the outer cold ($\sim$10$^4$~K) RSG or
LBV material.  As a result, a so-called conductive layer with
intermediate temperatures, $\sim$10$^5$~K, is created between the hot
and cool gas, as has been first suggested by \citet{Castor1975} and
\citet{Weaver1977} in wind-blown bubbles.  Observationally, the
conductive layer can be revealed by the presence of highly ionized
species \citep{Castor1975,Weaver1977}. Indeed, this conductive layer
has been detected in other wind-blown bubbles, such as those of
planetary nebulae \citep[PNe;][]{Ruiz2013} using far-UV \emph{FUSE}
observations of the O~{\sc vi} doublet.  At least in PNe, the
conductive layer at $\sim$10$^5$~K coexists with diffuse hot gas at
X-ray-emitting temperatures $\gtrsim$10$^6$~K \citep{Gruendl2004}.

The WR nebula around WR\,16 was first discovered by
\citet{Marston1994}, who described it as a round main nebula with
multiple arc-like features towards the northwest (see
Figure~\ref{fig:imagen_wr16}-left). The main nebula is classified as a
wind-blown bubble \citep[a $W$-type nebula according to the
classification scheme developed by][]{Chu1981}, whereas the outer
features have been associated with multiple mass ejection episodes
\citep{Marston1995}. The rim of the main nebula is sharper towards the
northwest direction. The diminished optical emission towards the
southeast may be associated with extinction due to the molecular
material along that direction that is revealed by radio observations
\citep{Marston1999,Duronea2013}. Alternatively, this morphology may
have resulted from the large proper motion of the central star along
this direction \citep{Duronea2013}.

The optical and radio morphologies of the nebula around WR\,16 are
consistent with the mid-infrared morphology revealed by \emph{WISE}
observations (see Figure~\ref{fig:imagen_wr16}-right).  In particular,
the W4 band (red in Figure~\ref{fig:imagen_wr16}-right) shows a
complete round shell with enhanced, limb-brightened emission towards
the northwest direction. The round WR nebula seen in this \emph{WISE}
band matches that of the \emph{IRAS} 60~$\mu$m image presented by
\citet{Marston1999}.

According to \citet{Toala2011}, the closed nebular shell and
limb-brightened morphology of a WR nebula make it a promising
candidate for its detection in X-rays, since a complete WR nebula is
able to confine the highly pressurized X-ray-emitting gas, as is the
case for S\,308 and NGC\,6888. Such hot gas can be expected to be in
contact with the cold nebular shell throughout a conductive layer.  In
this paper, we have searched for the different components of hot gas
in the WR bubble around WR\,16 using archival \emph{XMM-Newton} and
\emph{FUSE} observations towards this WR bubble. Section~2 describes
the observations and data processing, whereas we briefly describe our
findings in Section~3. We discuss our results in Section~4, and
finally we summarize them in Section~5.

\section{Observations}

\subsection{\emph{XMM-Newton} Observations}

The \emph{XMM-Newton} observations of the nebula around WR\,16 were
performed on 2009 Dec 28-29 (Obs.\ ID 0602020301; PI: S.\,Skinner)
during revolution 1841. The observations were obtained in the
framework of a survey to study single nitrogen-rich stars to determine
their X-ray properties and to characterize the emission processes.
The data have been presented by \citet{Skinner2012}, who focused on
the analysis of the X-ray emission from the central star, WR\,16.  No
further analysis of the potential diffuse X-ray emission was
performed, and thus we will concentrate in the search and analysis of
extended emission.

The two EPIC-MOS and pn cameras were operated in the full-frame mode
for a total exposure time of 33 and 30~ks, respectively.  
The medium optical blocking filter was used. 
The observations were processed using the \emph{XMM-Newton} Science Analysis 
Software (SAS Ver.\ 12.0.1) and the XMM-ESAS tasks for the analysis of the 
X-ray emission of extended objects and diffuse background \citep{Snowden2011}, 
as required for the analysis of \emph{XMM-Newton} observations of WR nebulae 
\citep{Toala2012}.  
The associated Current Calibration Files (CCF) available on June 2013, as obtained from {\small
  \url{ftp://xmm.esac.esa.int/pub/ccf/constituents/extras/esas_caldb}},
have been used to remove the contribution from the astrophysical
background, soft proton background, and solar wind charge-exchange
reactions with contributions at low energies ($<1.5$~keV).

\begin{figure*}[t]
\begin{center}
\includegraphics[width=0.53\linewidth]{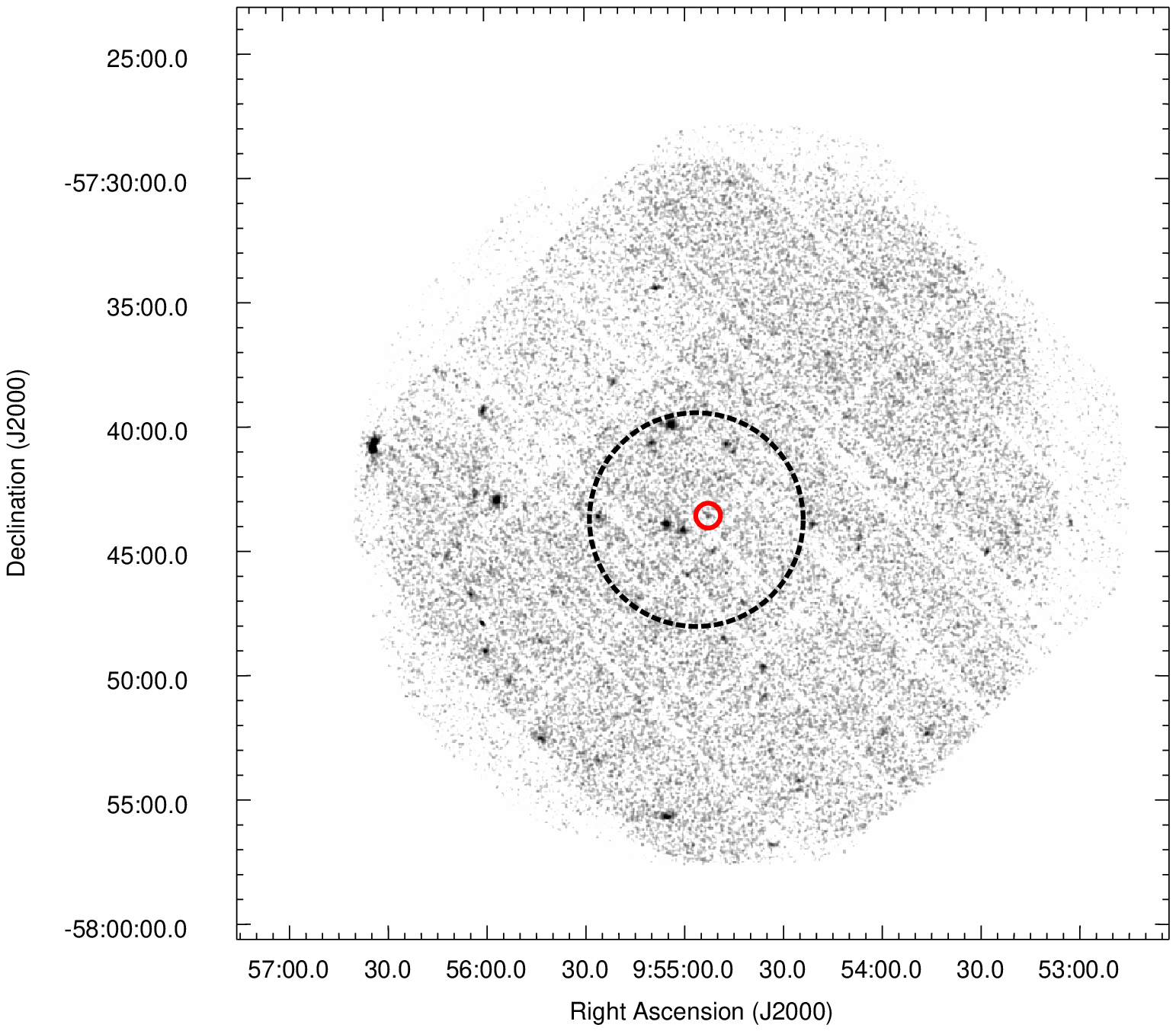}~
\includegraphics[width=0.445\linewidth]{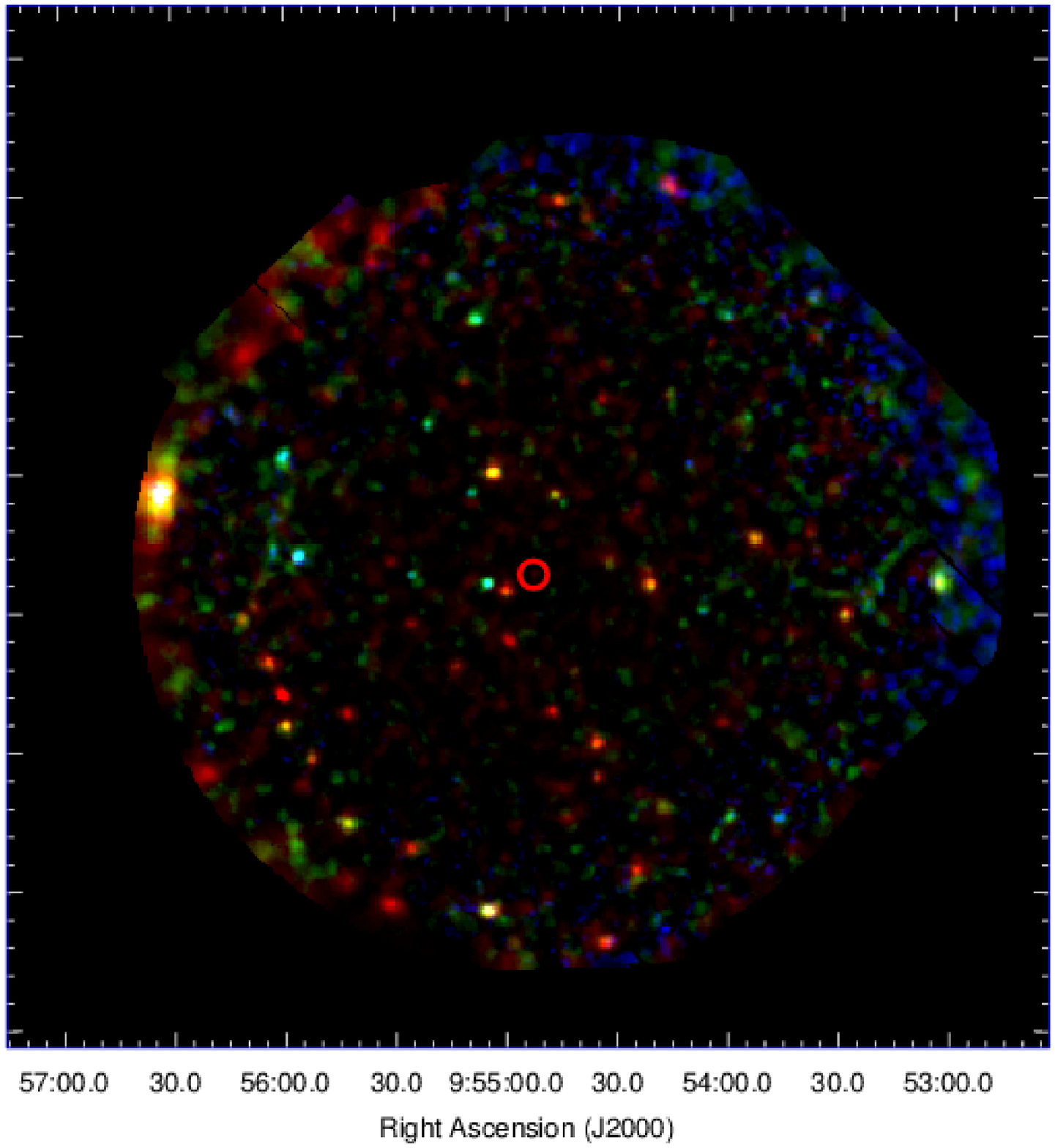}
\end{center}
\caption{ {\it (left)} \emph{XMM-Newton} EPIC (MOS1, MOS2, and pn)
  X-ray image of the field of view of WR\,16 for the total energy
  range 0.3--10 keV.  {\it (right)} \emph{XMM-Newton} EPIC smoothed
  exposure-corrected, color-composite X-ray picture of WR\,16, where
  the red color corresponds to the soft 0.3--1.15 keV energy band,
  green to the medium 1.15--2.5 keV band, and blue to the hard 2.5--10
  keV band.  The small red circle in both images shows the location of
  the central star, WR\,16, whereas the dashed-line circle in the left
  image encompasses the optical WR nebula.}
\label{fig:EPIC}
\end{figure*}

The net exposure times after processing the data are 27.5, 28.9,
and 14~ks for the EPIC/MOS1, MOS2, and pn cameras, respectively. 
We extracted images in four energy bands, namely 0.3-1.15 keV (soft), 
1.15-2.5 keV (medium), 2.5-10 keV (hard), and 0.3-10 keV (total).  
The combined EPIC X-ray image in the total energy band is shown in 
Figure~\ref{fig:EPIC}-{\it left}.  
Smoothed exposure-corrected images of the soft, medium, and hard images were 
created using the ESAS task \emph{adapt-900} requesting 20 counts of the 
original image for each smoothed pixel, and these were combined to create 
the false-color picture shown in Figure~\ref{fig:EPIC}-{\it right}.

\subsection{\emph{FUSE} Observations}

To complement the X-ray observations towards the WR bubble around
WR\,16, we have searched \emph{Far Ultraviolet Spectroscopic Explorer}
(\emph{FUSE}) observations on the Mikulski Archive for Space
Telescopes (MAST) of WR\,16. The \emph{FUSE} observations have a
spectral coverage between 920--1190\AA\, with spectral resolution of
R$\sim20\,000$ \citep{Moos2000,Sahnow2000}.  Two sets of observations
were used to create a far-UV spectrum of WR\,16, namely Obs.\ ID
G9271501 (start time 2006-07-10 05:18:34; exposure time 30.6~ks) and
G9271502 (start time 2006-07-11 12:34:37; exposure time 30.7~ks).  The
data were reprocessed with the CalFUSE calibration pipeline software
package, CalFUSE 3.2.3 \citep{Dixon2007}.

The resultant spectrum will be used to analyze the O~{\sc vi}
$\lambda\lambda$1032,1037 doublet.  The presence of these lines may
imply a conductive layer of gas at $T\sim$3$\times$10$^5$~K between
the outer cold (10$^{4}$~K) nebular material and the interior hot
content, as in the case of star forming regions
\citep[e.g.][]{Pathak2011} and PNe \citep{Ruiz2013}. The O~{\sc vi}
lines are registered by the 
LiF2B (979--1075 \AA), and 
\AA) and SiC2B (1016--1103 \AA) telescopes, gratings, and detector
segments combinations.  After processing, the observation ID G9271501
has net exposure times of 28.4 ks for LiF1A and 30.5 ks for LiF2B,
whereas the observation ID G9271502 has net exposure times of 25.2 ks
for LiF1A, 26.5 ks for LiF2B, and 7.2 ks for SiC1A.  For both
observations, the quality of the spectrum registered by SiC2B was
limited and the resulting spectrum was not used. Details of the data
processing and merging are similar as those described by
\citet{Guerrero2013}.

\begin{figure*}[t]
\begin{center}
\includegraphics[width=0.825\linewidth]{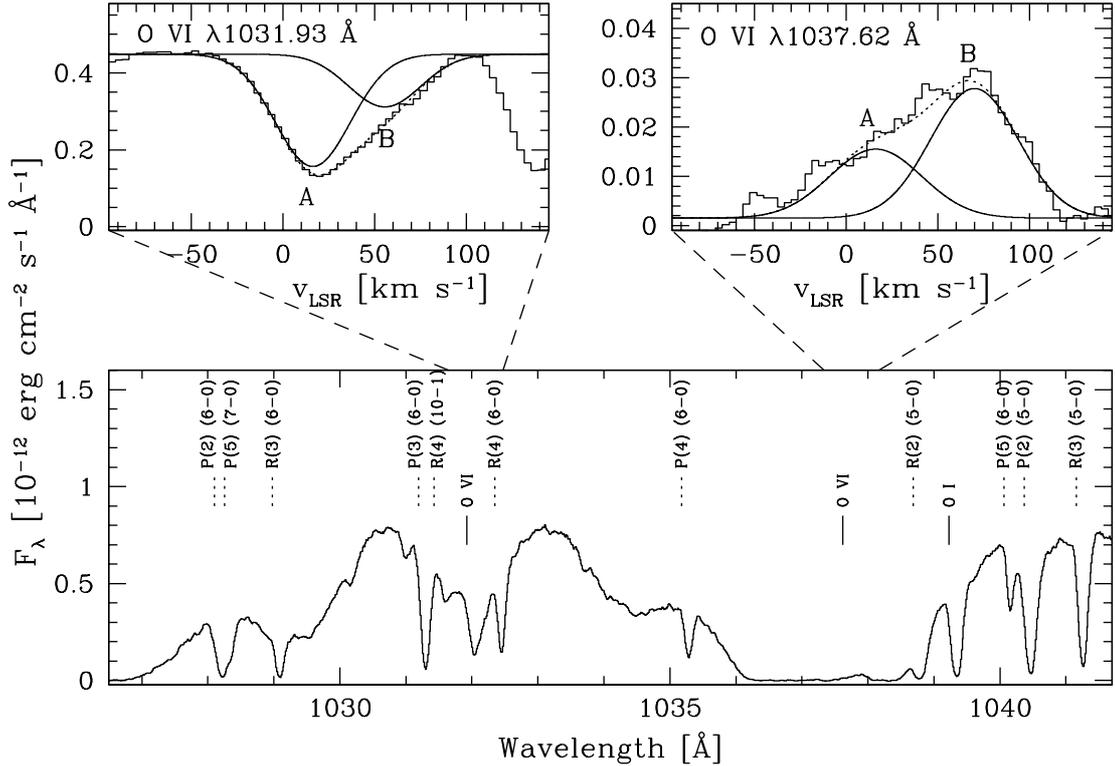}
\end{center}
\caption{ \emph{FUSE} far-UV spectrum towards WR\,16.  The bottom
  panel shows the spectral range around the O~{\sc vi} lines.
  Vertical solid lines mark the position of the O\,{\sc vi} doublet
  and O~{\sc i} atomic line, whereas dotted lines mark the position of
  H$_2$ absorption lines.  The upper panels show enlargements of the
  O~{\sc vi} lines in the LSR velocity frame with multiple Gaussian
  fits to their profiles, which include components A and B as
  described in the text.}
\label{fig:OVI}
\end{figure*}

Figure~\ref{fig:OVI} shows the final spectrum obtained from the
\emph{FUSE} observations. 
The O~{\sc vi} doublet lines are marked at 1031.93 \AA\ and 1037.62 \AA. 
Additional absorption lines from material in the interstellar medium 
and Earth atmosphere are also shown. 
Details and discussion on this spectrum will be presented in Section~4.

\section{Results}

A preliminary inspection of the images presented in
Figure~\ref{fig:EPIC} does not reveal any diffuse X-ray emission
within the nebula around WR\,16, although a significant number of
point sources are detected in the field of view of the EPIC cameras.
Even though there is no detection of diffuse X-ray emission within
this WR bubble, we can estimate an upper limit from the present
observations.  To derive this upper limit, we have extracted
background-subtracted spectra from the three EPIC instruments
corresponding to the source region shown in
Figure~\ref{fig:EPIC}-left, which encloses the optical nebula with a
radius of 4\farcm3.  Point sources were carefully excised from either
the source and background regions.  The spectral analysis allowed us
to derive 3-$\sigma$ upper limits to the EPIC-pn, EPIC-MOS1, and
EPIC-MOS2 count rates in the 0.3--1.5~keV energy range $<$28
counts~ks$^{-1}$, $<$14~counts~ks$^{-1}$, and $<$13~counts~ks$^{-1}$,
respectively.

The X-ray flux and luminosity upper limits for the WR\,16 nebula can
be estimated from the count rates given above using the count rate
simulator WebPIMMS (version 4.6)\footnote{{\small
    \url{heasarc.gsfc.nasa.gov/Tools/w3pimms.html}}}. To derive these
values, we have adopted an optically-thin APEC plasma emission model
absorbed by a column density of 4.1$\times$10$^{21}$~cm$^{-2}$, as
implied by Bohlin et al.'s (1978) relation for an optical extinction
$A_{\mathrm{V}}$ of 2.14~mag where the average of the values is given
by \citet{vdH2001}.  Furthermore, the temperature of the
X-ray-emitting gas in the nebula has been assumed to be
$\sim$1.4$\times$10$^6$ K ($kT\sim0.122$ keV), i.e., an intermediate
value between those reported for the hot gas detected in S\,308 and
NGC\,6888. Accounting for the use of the EPIC-pn camera with the
medium filter, the resultant 3-$\sigma$ upper limit for the absorbed
flux in the 0.3--1.5~keV energy band is
$f_{\mathrm{X}}\lesssim$3.1$\times$10$^{-14}$ erg~cm$^{-2}$~s$^{-1}$
for an intrinsic flux of $F_{\mathrm{X}}\lesssim$1.1$\times$10$^{-12}$
erg~cm$^{-2}$~s$^{-1}$. The corresponding upper limit for the X-ray
luminosity is
$L_{\mathrm{X}}\lesssim$7.4$\times$10$^{32}$~erg~s$^{-1}$ for a
distance of $d=2.37$~kpc \citep{vdH2001}.


Using the normalization factor\footnote{$A=$1$\times10^{-14} \int
  n_{\mathrm{e}} dV/4 \pi d^{2}$, where $d$ is the distance,
  $n_{\mathrm{e}}$ the electron density, and $V$ the volume.} given by
WebPIMMS, $A=$1.38$\times$10$^{-3}$~cm$^{-5}$, we can estimate an
upper limit to the electron density of the hot gas in the nebula. This
gives $n_{\mathrm{e}}<$0.6~cm$^{-3}$.

The \emph{FUSE} spectrum shows a strong absorption feature at the
wavelength of the O~{\sc vi} $\lambda$1031.93 \AA\ line and one weak
emission feature at the wavelength of the O~{\sc vi} $\lambda$1037.62
\AA. These are indicative of the presence of hot gas at
$\sim$3$\times$10$^5$ K along the line of sight of WR\,16. The
detection of one component of the O~{\sc vi} doublet in absorption and
the other in emission is certainly puzzling, as the physical
conditions for emission and absorption are very different: O~{\sc vi}
emission is typically ascribed to narrow, high-density interfaces in
the local interstellar medium \citep{DS08}, whereas O~{\sc vi}
absorptions can also be produced in large volumes of low-density hot
gas.  The two lines have been reported to be detected either in
emission or absorption in surveys of the Galaxy and Magellanic Clouds
\citep[e.g.,][]{Howk_etal02}, but the simultaneous detection of one
component in absorption and the other in emission requires singular
observational conditions. This may be the case of WR\,16, as the
stellar continuum around the O~{\sc vi} $\lambda$1037.62 \AA\ line is
completely saturated, whereas it is not around the O~{\sc vi}
$\lambda$1031.93 \AA\ component. The particular shape of the stellar
continuum of WR\,16 makes the O~{\sc vi} $\lambda$1037.62 \AA\
component sensitive to emission from material along the line of sight
but insensitively to absorption.  On the other hand, the O~{\sc vi}
$\lambda$1031.93 \AA\ component is sensitive both to emissions and
absorptions along the line of sight, but the former dominates the
profile of this line, as the absorption component is stronger than the
emission.

The profile shape of the O~{\sc vi} lines is not symmetric, implying
different velocity components. We find that the most simple fit of
both the emission and absorption spectral features involves two
Gaussian components with $V_{\mathrm{LSR}}\simeq$16 km~s$^{-1}$ and 60
km~s$^{-1}$, which are marked as component A and B in the top panels
of Figure~\ref{fig:OVI}, respectively.  Component A has the same
radial velocity as the H$_2$ and O~{\sc i} absorption lines
overimposed in the spectrum of WR\,16, which are visible at the bottom
panel of Figure~\ref{fig:OVI}. Following the measurement of equivalent
widths and column densities technique of O~{\sc vi}
\citep{Savage1991}, we have computed the column density of the O~{\sc
  vi} $\lambda$1031.93 \AA\ line, $N_{\mathrm{OVI}}$ (in cm$^{-2}$),
along the line of sight of WR\,16.  This technique uses the apparent
optical depth in terms of velocity,
\begin{equation}
  \tau(v) = \ln[I_{\mathrm{o}}(v)/ I_{\mathrm{obs}}(v)],
\end{equation}
where $I_{\mathrm{o}}(v)$ and $I_{\mathrm{obs}}(v)$ are the continuum
intensity and absorption line depth, respectively.  The parameter
$N_{\mathrm{OVI}}$ can be calculated as
\begin{equation}
  N_{\mathrm{OVI}}=\frac{m_{\mathrm{e}}c \tau(v)}{\pi \mathrm{e}^{2}f \lambda}\times \Delta V_{\mathrm{FWHM}}=3.768\times10^{14}\frac{\tau(v)}{f \lambda}\Delta V_{\mathrm{FWHM}},
\end{equation}
where $\lambda$ is the wavelength in \AA and $f$ is the oscillator
strength of the atomic species \citep[for O~{\sc vi}, $f$=0.1325;
see][for details]{Pathak2011}. The parameters $c$, $m_{\mathrm{e}}$,
and $\mathrm{e}$ are the speed of light, and electron mass, and
charge, respectively, where $\Delta V_{\mathrm{FWHM}}$ is the full
width at half maximum of the line in km~s$^{-1}$. The O~{\sc vi}
column density is estimated to be 1.4$\times$10$^{14}$ cm$^{-2}$,
which is consistent with the averaged value expected for a star at
2.37 kpc
\citep[$N_{\mathrm{OVI}}\lesssim$10$^{14}$~cm$^{-2}$;][]{deAvillez2012}.
This is especially true if we consider the large spatial variations
revealed by studies of the O~{\sc vi} content in the local
interstellar medium along the line of sight of B-stars \citep{WL08}
and hot DA white dwarfs \citep{Barstow_etal10}. Therefore, we can
conclude that component A is due to the interstellar medium along the
line of sight of WR\,16.

\begin{table*}[t]
\label{tab:nebulosas}
\centering
\caption{Stellar and nebular parameters for the WR nebulae observed with modern X-ray satellites.}
\begin{tabular}{llcccccccc}
\hline
\hline
Star    & WR Nebula & Distance$^{\mathrm{a}}$  & WR type$^{\mathrm{a}}$ &$v_{\infty}^{\mathrm{a}}$  & \multicolumn{2}{c}{Radius$^{\mathrm{b}}$}   & $N_{\mathrm{H}}^{\mathrm{c}}$  &S$^{\mathrm{c}}$        & $L_{\mathrm{X}}^{\mathrm{c}}$       \\ \cline{6-7}
        &           & (kpc)     &     & (km~s$^{-1}$)& (arcmin)  & (pc) & ($\times10^{21}$~cm$^{-2}$)  &(erg~cm$^{-2}$~s$^{-1}$~arcmin$^{-2}$)  &($\times10^{33}$~erg~s$^{-1}$)  \\
\hline
WR\,6   & S\,308    & 1.5       & WN4 & 1800       & 20      & 8.8  & 0.62--1.1      &  3.5$\times$10$^{-15}$    & 2             \\
WR\,16  &           & 2.37      & WN8h& 630        & 4.3     & 3    & 4.1            &  $<$5.4$\times$10$^{-16}$ &$<$0.8         \\
WR\,40  & RCW\,58   & 2.26      & WN8h& 840        & 4.9     & 3.2  & 5              &  $<$2.2$\times$10$^{-15}$ &$<$6.5         \\
WR\,136 & NGC\,6888 &1.26       & WN6 & 1750       & 8.6     & 3.2  & 3.1            &  1.2$\times$10$^{-14}$    & 3.5           \\
\hline
\hline
\end{tabular}
\begin{list}{}{}
\item{
$^{\mathrm{a}}$Distance, spectral type, and stellar wind velocity taken 
from \citet{vdH2001}, except for the distance of WR\,6 that was taken 
from the kinematic distance estimated by \citet{Chu2003}.}
\item{
$^{\mathrm{b}}$For NGC\,6888 this is the semi-major axis.}
\item{ $^{\mathrm{c}}$Column density taken from \citet{Toala2012} for
  S\,308, \citet{Gosset2005} for RCW\,58, and \citet{Zhekov2011} for
  NGC\,6888.  The observed X-ray surface brightness and luminosity of
  these sources have been scaled from these references to the
  distances used here and extended to the 0.3--1.5 keV energy range in
  all cases for consistency.}
\end{list}
\end{table*}

Component B of the O~{\sc vi} doublet, on the other hand, does not
have a molecular or neutral interstellar medium counterpart.  The
intensity of its absorption in the $\lambda$1031.93 \AA\ line implies
a column density of 4.5$\times$10$^{13}$ cm$^{-2}$, which is, about
one third that of component A.  The conductive layer within the nebula
would result in a narrow O~{\sc vi} absorption feature bluewards of
the systemic velocity of the WR\,16 nebula.  There are, however, no
measurements of the dynamics (radial velocity and expansion) of the
optical main WR bubble around WR\,16 to carry out such comparison.
\citet{Duronea2013} attributed a molecular component at
$V_\mathrm{LSR}\simeq-9$ km~s$^{-1}$ to the main nebular shell around
WR\,16.  If this were indeed the case, then component B could not be
ascribed to hot gas inside the WR nebula around WR\,16 as its velocity
would be redwards of the systemic velocity.  We should then conclude
that component B is most likely associated with a high excitation
component of the interstellar medium along the line of sight towards
WR\,16.

\section{Discussion}

The WR nebulae can be expected to be filled with hot diffuse gas, as
the current WR wind slams the slow and dense, previously ejected RSG
or LBV material, which creates a shock wave that thermalizes the
interior of the WR nebula. The star WR\,16 possesses a stellar wind
with terminal velocity, $v_{\infty}$, of 630~km~s$^{-1}$
\citep{vdH2001}, which would imply a shocked gas temperature
$\sim$1.2$\times$10$^7$ K \citep[for a mean free particle
$\mu\sim1.3$;][]{vanderHucht1986}. Higher velocities would imply even
higher plasma temperatures, but S\,308 and NGC\,6888 possess plasma
with temperatures barely greater than $10^{6}$~K, which contrasts the
terminal wind velocities, $\gtrsim$1700 km~s$^{-1}$, of their central
stars. This discrepancy is also seen in {\sc H ii} regions and PNe, in
which thermal conduction is invoked to explain the low observed plasma
temperatures of the shocked stellar wind \citep{Chu2008}.

The \emph{XMM-Newton} observations described here show that no
extended X-ray emission is detected inside the main nebular shell of
WR\,16, despite the sensitive integration time of these observations.
To put this non-detection in context with previous observations of WR
nebulae, we show the stellar and nebular parameters (distance,
spectral type of the WR star, stellar wind velocity, nebular radius,
observed X-ray surface brightness, and luminosity) for the four WR
nebulae observed to date in Table~1 by the present generations of
X-ray satellites, namely S\,308, NGC\,6888, RCW\,58, and that around
WR\,16. The surface brightness and luminosity of RCW\,58 in the
0.3--1.5~keV energy range have been computed from the count rate
reported by \citet{Gosset2005} for a distance of 2.26~kpc
\citep{vdH2001} after adopting the same plasma temperature as for the
WR nebula around WR\,16. The observed surface brightness presented in
Table~1, which is obtained by dividing the observed X-ray flux by the
area sustained by the nebula, provides a distance independent
measurement of the relative X-ray brightness of each nebula.  We
remark that the X-ray surface brightness and luminosity for RCW\,58
and the WR nebula around WR\,16 given in this table correspond to
upper limits.

We first note that the distances to all WR nebulae in Table~1 are not
dramatically different, ranging from 1.26 kpc up to 2.37 kpc.
Furthermore, they are all absorbed by similar hydrogen column
densities in the range 6$\times$10$^{20}$--5$\times$10$^{21}$
cm$^{-2}$. This situation suggests that the detection of extended
X-ray emission from these sources does not depend on external factors,
such as distance or amount of intervening material but rather points
that the nebula around WR\,16 is intrinsically fainter than S\,308 and
NGC\,6888 with a surface brightness 6--20 times lower.

In sharp contrast, the stellar properties (spectral type and terminal
wind velocity) seem to be correlated with the detectability of
extended X-ray emission within these WR nebulae. Both S\,308 and
NGC\,6888 harbor early WN4 and WN6 stars, respectively, with terminal
wind velocities $\gtrsim$1700 km~s$^{-1}$. Meanwhile, the non-detected
WR nebulae in the X-rays have late WN stars, WN8h, with stellar
velocities below 900~km~s$^{-1}$.

The nebular morphology has been claimed in the past to be correlated
with X-ray emission: diffuse X-ray emission is present in S\,308 and
NGC\,6888, and possess undisrupted morphologies, whereas RCW\,58,
which is undetected in X-rays, has a disrupted morphology
\citep{Gruendl2000}. The interpretation of this behaviour is that a
disrupted shell cannot hold the hot gas which escapes away.  Contrary
to this trend, the nebula around WR\,16 displays a complete shell
morphology (Figure~\ref{fig:imagen_wr16}), but we have set a stringent
upper limit for the X-ray emission from hot plasma in its interior.

The lack of X-ray emission motivated us to investigate the \emph{FUSE}
spectrum of WR\,16, to search for spectral features in the O~{\sc vi}
lines that could be attributed to gas at temperatures of
$\sim$3$\times$10$^5$ K (Figure~\ref{fig:OVI}). Although several
O~{\sc vi} absorption and emission components are detected in the
\emph{FUSE} spectrum of WR\,16, none of them can be unambiguously
attributed to a conductive layer in its nebula.  This seems to suggest
that this bubble has not experienced the effects of thermal conduction
or any other equivalent physical processes \citep[e.g., hydrodynamic
ablation, photoevaporation;][]{Arthur2007,Pittard2007} in the
wind-wind interaction zone.  As a result, the shocked wind material
may still possess temperatures of 1.2$\times$10$^7$~K but extremely
low densities in the range 0.001--0.01~cm$^{-3}$. If this would be the
case, the non-detection of X-ray emission from the WR bubble around
WR\,16 does not automatically imply the lack of interior hot gas, but
its low density has reduced its differential emission measure below
the detectability limits.

\section{Summary and conclusions}

We present the analysis of \emph{XMM-Newton} archival observations of
the WR nebula around WR\,16.  These observations show that there is no
detection of hot gas within the nebula, but a number of point sources
are detected in the field of view.  An upper limit to the unabsorbed
flux in the 0.3--1.5~keV band has been estimated to be
$F_{\mathrm{X}}\lesssim$1.1$\times$10$^{-12}$~erg~s$^{-1}$~cm$^{-2}$,
which corresponds to a luminosity
$L_{\mathrm{X}}\lesssim$7.4$\times$10$^{32}$~erg~s$^{-1}$ at a
distance of 2.37~kpc and an upper limit to the electron density of
$n_{\mathrm{e}}<$0.6~cm$^{-3}$.

Following the hypothesis that thermal conduction takes place between
the outer cold (10$^4$~K) nebular gas and the hot interior of the WR
nebula, we have searched for a conductive layer between these two
components using \emph{FUSE} observations of the O~{\sc vi} doublet.
We find that both lines can be model as the contribution of two
components at $V_{\mathrm{LSR}}$=16~km~s$^{-1}$ and
$V_{\mathrm{LSR}}$=60 km~s$^{-1}$, respectively. The velocity of the
first component is coincident with that of H$_2$ and low-excitation
ions and neutral atoms in the interstellar medium along the line of
sight towards WR\,16.  Its column density is also consistent with the
amount of intervening material expected for a star at the distance of
WR\,16.  The second component does not have a counterpart in the
low-excitation interstellar medium, but its radial velocity is not
consistent with that expected for the expanding nebular shell of
WR\,16 from radio molecular observations. Unless there is a
significant revision of the radial velocity of WR\,16, the present
observations suggest that this second component can neither be
associated with a conductive layer.

The lack of detection of hot gas (10$^5$--10$^7$ K) within the WR
bubble around WR\,16 is puzzling. This nebula is one of the four WR
nebulae observed with the latest generation of X-ray satellites, and
is the second non-detected. Because the low number statistics, it is
important to increase the number of X-ray observations towards these
nebulae with better spatial coverage, sensitivity, and energy
resolution to define the stellar and nebular properties required for
the presence of hot gas in these objects.

\acknowledgements We would like to thank the anonymous referee for
her/his valuable comments, which improved the presentation of this
paper.

Part of the data presented in this paper were obtained from the Mikulski 
Archive for Space Telescopes (MAST). 
STScI is operated by the Association of Universities for Research in 
Astronomy, Inc., under NASA contract NAS5-26555. 
Support for MAST for non-HST data is provided by the NASA Office of 
Space Science via grant NNX09AF08G and by other grants and contracts.  

This work is funded by the Spanish MICINN (Ministerio de Ciencia e 
Innovaci\'on) grant AYA 2011-29754-C03-02 co-funded with FEDER funds.
JAT acknowledges support by the CSIC JAE-Pre student grant 2011-00189.

\end{document}